\documentstyle[prd,aps,epsfig,preprint]{revtex}
\begin{document}

\preprint{                                                 BARI-TH/280-97}
\draft
\title{          Upward-going muons and neutrino oscillations            }
\author{         G.~L.~Fogli, E.~Lisi, and A.~Marrone}
\address{   Dipartimento di Fisica and Sezione INFN di Bari, \\
                  Via Amendola 173, I-70126 Bari, Italy}
\maketitle
\begin{abstract}
The available upward-going muon data from the Kamiokande, Baksan, MACRO, 
IMB, and SuperKamiokande experiments are reviewed and combined. Bounds on 
the neutrino mass and mixing parameters are derived for oscillations in 
two and three flavors. These bounds are not in significant conflict with 
the oscillation solution to the atmospheric neutrino flavor anomaly observed
in the sub-GeV and multi-GeV energy range. The combination of all the 
available atmospheric data tends to favor the $\nu_\mu\leftrightarrow\nu_e$  
channel with respect to the $\nu_\mu\leftrightarrow\nu_\tau$ channel, and to 
disfavor the threefold maximal mixing scenario.
\end{abstract}
\pacs{\\ PACS number(s): 96.40.Tv, 14.60.Ef, 25.30.Pt, 14.60.Pq}

\section{Introduction}

	The observation of upward-going muons in underground detectors has
long since been considered \cite{Ma61} as an effective tool to study 
the neutrino component expected in atmospheric showers \cite{Le63,Os65,Co66}. 
Early anomalies in the muon event samples detected by the first two 
dedicated experiments \cite{Joha,Kola} soon sparked investigations of 
possible neutrino oscillation effects \cite{Sa76,Br80}.

	The upward-going muon total rates measured in recent underground
experiments appear to be approximately in agreement with the expectations, 
thus disfavoring large neutrino oscillation effects. This contrasts with 
the observed anomaly in the electron and muon contents of (semi)contained 
events induced by atmospheric neutrinos, that might signal flavor oscillation 
effects (see \cite{Ga96} for a recent review). The conflict between these 
two data sets is usually resolved by appealing to the large, noncancelable 
uncertainties affecting the calculation of the {\em absolute\/} muon rates 
\cite{Ga97,St95,Ho95}.%
\footnote{
We add that the experimental systematics might also be more significant
than usually assumed, see Sec.~II~B.
}
In terms of neutrino oscillations, this  implies that there are values
of the oscillation parameters that explain the ``anomalous'' muon-to-electron 
flavor ratio  without really conflicting with the ``regular'' upward muon 
data \cite{Fr93,Ak93,Li95}. In this work we confirm and elucidate such point 
of view by performing  an updated analysis of the 
upward-going muon data from five experiments: 
Kamiokande \cite{Oy89,Mo91,Mo93,Sz96}, 
Baksan \cite{Bo95,Mi96,Su96},
MACRO \cite{Ah95,Ro96}, 
IMB \cite{Ca91,Be92}, and 
SuperKamiokande \cite{Yo97,Su97}.

	The paper is structured as follows. In Sec.~II the available data 
on upward-going muons are reviewed, compared  with each other and with the
theoretical expectations (in the absence of oscillations), and then combined.
In Sec.~III the combined data on the muon flux angular distribution are used 
to derive  bounds on the two-flavor and three-flavor neutrino oscillation 
parameters. These bounds are compatible with the neutrino oscillation solution 
to the atmospheric neutrino anomaly. It is shown that the inclusion of 
upward-going muon data in fits tends to favor the 
$\nu_\mu\leftrightarrow\nu_e$ oscillation channel and to disfavor the 
threefold maximal mixing scenario. In Sec.~IV we draw the conclusions of 
our work. This study is part of a vast research program aiming to analyze 
the world neutrino oscillation data, including the results of solar 
\cite{Fo96}, atmospheric, \cite{Fo96,Fo95,Fo97} and laboratory 
(accelerator and reactor) \cite{Fo96,Sc95}  neutrino experiments.

\section{Experimental Data and Standard Expectations}

	The experimental data on upward-going muons are usually given
as distributions of either muon events or muon fluxes as a function of the 
zenith angle $\theta$, with $\cos\theta=0$ $(\cos\theta=-1)$ corresponding to 
horizontal (vertical) muons. The distribution of {\em muon events\/} depends  
upon technical specifications of the detector such as the efficiency and the 
geometric acceptance, that are often unpublished. The distribution
of {\em muon fluxes\/} (when available) is instead deconvoluted 
for these effects, and thus allows a more direct comparison with
the theoretical expectations. We describe in Sec.~II~A  the more recent 
available data (either  muon events or muon fluxes) and in Sec.~II~B the 
specific combination of data used in our oscillation analysis.

\subsection{Data from different experiments}

\begin{figure}
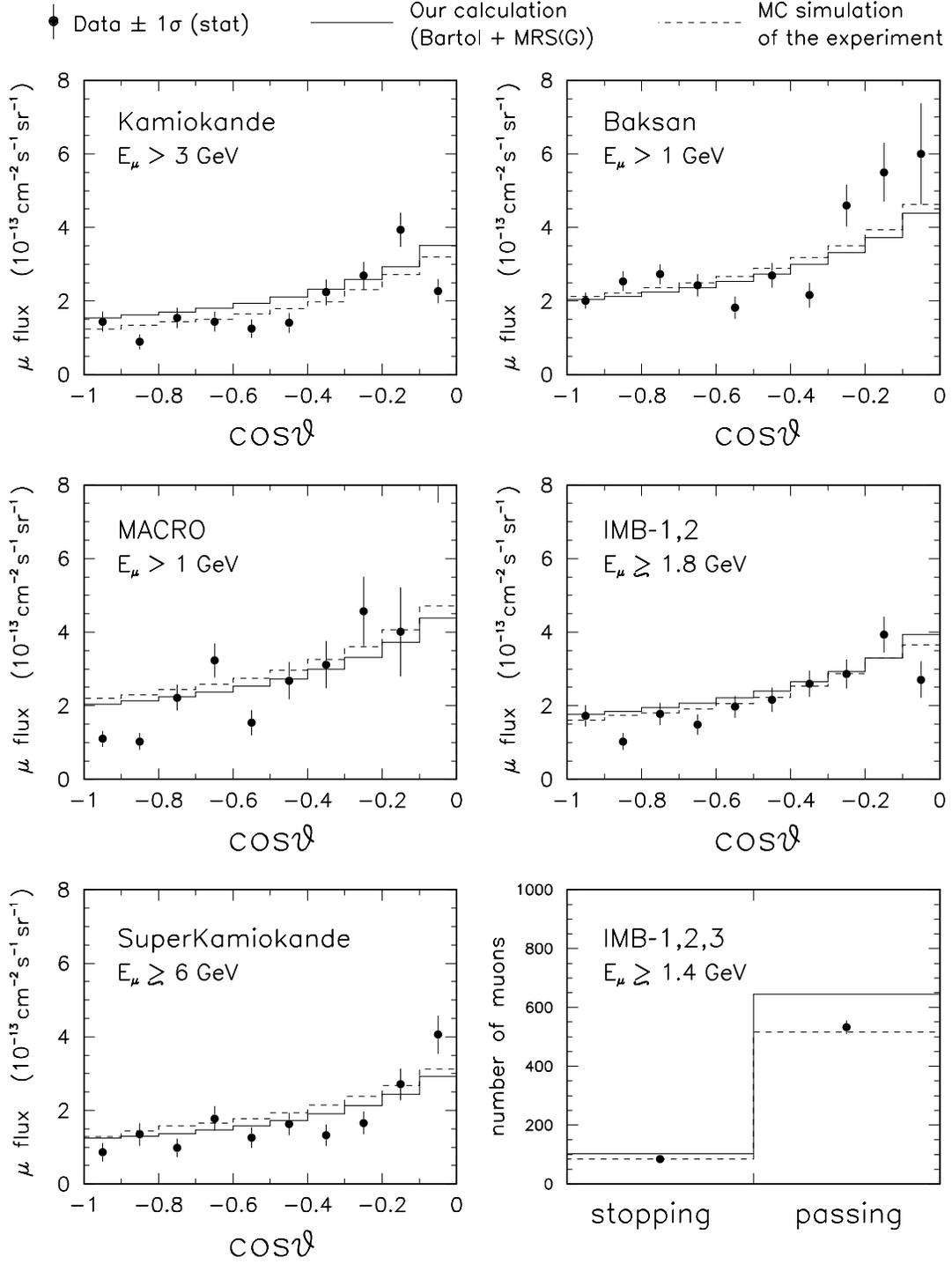

\caption{	Compilation of upward-going muon data and expectations
		for different experiments: 
		Kamiokande \protect\cite{Sz96},
		Baksan \protect\cite{Mi96,Su96},
		MACRO \protect\cite{Ro96},
		SuperKamiokande \protect\cite{Su97}
		IMB-1,2 \protect\cite{Ca91},
		and IMB-1,2,3 \protect\cite{Be92}. Dots with
		error bars: Experimental data with $1\sigma$
		statistical uncertainties. Solid histograms:
		our calculations. Dashed histograms: Experimental
		simulations. $\theta$ is the zenith angle. See also
		Table~I.}
\label{fig:1}
\end{figure}

	In the last few years, data on upward going muons have
been collected by five experiments: 
	Kamiokande 		\cite{Oy89,Mo91,Mo93,Sz96}, 
	Baksan 			\cite{Bo95,Mi96,Su96}, 
	MACRO 			\cite{Ah95,Ro96}, 
	IMB 			\cite{Ca91,Be92}, and
	SuperKamiokande 	\cite{Yo97,Su97}. 
The IMB experiment (completed) was performed in three distinct stages 
(phases 1, 2, and 3); we will comment both on the global results (IMB-1,2,3)
\cite{Be92} and on the results from the first two phases (IMB-1,2) 
\cite{Ca91}. Data from the pioneering NUSEX experiment \cite{Ag91}
(eight events only) will not be considered. Since some experimental results 
have been reported only in conference proceedings, we compile the most recent
data on upward-going muons for the benefit of the reader. The compilation 
is reported in Table~\ref{tab:1} and in Fig.~\ref{fig:1}, that we comment 
in parallel.

	Table~\ref{tab:1} reports, for each experiment, the relevant
detector characteristics and the experimental muon data, compared with the 
corresponding expectations in ten angular bins. The theoretical muon fluxes 
refer to our calculation%
\footnote{
A general description of the theoretical estimate of muon fluxes
can be found, e.g., in \cite{Ga84,Au88}.
}
with the following inputs: Bartol neutrino fluxes 
\cite{Ag96} (neglecting small, site-dependent geomagnetic effects), version 
``G'' of the Martin-Stirling-Roberts [MRS(G)] structure functions \cite{MRSG}
(i.e., the default choice of the CERN Parton Density Function Library 
\cite{PDFL,PB95}), and Lohmann-Kopp-Voss (LKV) muon energy losses in standard 
rock \cite{Lo85}. The use of the same input for the calculation of the 
expected muon fluxes will enable us  to perform a uniform comparison with 
all the experiments.%
\footnote{It should be stressed, however, that alternative input choices
for the neutrino fluxes or for the structure functions may induce 
variations as large as $\sim30\%$ in the normalization of the expected 
muon fluxes. This uncertainty will be included in the fits.}
\ Notice that, for fixed inputs, the theoretical muon spectrum depends 
only on the energy threshold; therefore, the Baksan and MACRO experiments
(both with $E_\mu>1$ GeV) share the same expectations.
All other experiments have higher thresholds%
\footnote{In particular, the IMB threshold corresponds  to $\sim1.8$ GeV
for phases 1 and 2, to $\sim1$ GeV for phase 3, and to an effective value 
$\sim 1.4$ GeV for the total data sample (IMB-1,2,3).}
 and correspondingly lower muon fluxes. Concerning the experimental data, 
we quote only the statistical errors, since the systematic errors in each bin
are usually unpublished. The only notable exception is the MACRO analysis 
\cite{Ah95,Ro96}, where the systematic uncertainties are estimated to range
from a few percent near the vertical to about $20\%$ near the horizontal
direction. We also report (last row in Table~\ref{tab:1}) 
the total muon flux, defined 
as the sum of the individual fluxes in each bin times the bin width,
with (statistical) errors added in quadrature.

	Figure~\ref{fig:1} displays, in graphical form, the same information as
Table~\ref{tab:1}. 
The data are shown as dots with (statistical) error bars. Our 
calculations (in the absence of oscillations) are represented by the solid 
histograms.  In addition,  we report as dashed histograms
the original MonteCarlo simulations of the various collaborations. 
The ingredients of these simulations were:
	(1) Kamiokande \cite{Oy89,Mo93}: 
Volkova neutrino fluxes \cite{Vo80}, 
Eichten-Hinchliffe-Lane-Quigg (EHLQ) structure functions \cite{EHLQ,PDFL}, and 
Lohmann-Kopp-Voss (LKV) muon energy losses \cite{Lo85}; 
	(2) Baksan \cite{Mi96,Su96}: 
Bartol neutrino fluxes \cite{Ag96},
Morfin-Tung (MT) structure functions \cite{MoTu}, and 
LKV muon energy losses; 
	(3) MACRO \cite{Ro96}: 
Bartol neutrino fluxes,
MT structure functions, and 
LKV muon energy losses; 
	(4) IMB \cite{Ca91,Be92}: 
Volkova neutrino fluxes,
EHLQ structure functions, and 
Bezrukov-Bugaev muon energy losses \cite{Be81}; 
	(5) SuperKamiokande \cite{Su97}: 
Bartol neutrino fluxes, 
EHLQ structure functions, and 
LKV muon energy losses.
The differences between our calculations and the MonteCarlo
simulations are relatively small.

The last panel in Fig.~\ref{fig:1} shows the total number of muons that stopped
in, or passed through, the IMB detector (phases 1, 2, and 3 combined).%
\footnote{
The muon flux distribution for IMB-1,2,3 is, unfortunately, unpublished.
}
 Unfortunately, it is difficult to calculate reliably the
corresponding theoretical rates without detailed information about
the IMB detector. The very definition of ``passing'' or ``stopping'' muon 
depends on unreproducible geometric cuts applied on an event-by-event basis.
Their approximation with (more manageable) energy cuts (see, e.g., 
\cite{Ak93}) may bias  the classification of events.  Moreover, the absolute 
estimates of muon yields requires also the accurate knowledge
of the  detector geometrical acceptance and of its energy-dependent
efficiency, which are not published. Therefore,  we are not surprised that, 
just for IMB-1,2,3 (last panel of Fig.~\ref{fig:1}), our  tentative calculation of the 
``passing'' muon yield is not in good agreement with the published IMB 
MonteCarlo simulation \cite{Be92}. For these reasons, we have not used the 
IMB-1,2,3 stopping and passing muon data in our analysis. We have instead used
the muon flux distribution (published for IMB-1,2 \cite{Ca91}), that can be 
calculated more reliably and can also be compared with similar 
distributions from other experiments.

\subsection{Combination of data}

	As Fig.~\ref{fig:1} shows, the angular distributions measured by
Kamiokande, Baksan, MACRO, IMB-1,2, and SuperKamiokande
are  roughly in agreement with the theoretical expectations.
However, the data also show some  ``bumps'' or ``dips'' that seem too 
large to be statistical fluctuations but also too random to suggest new 
physics---they are likely to be the effect of unknown experimental 
systematics. Let us consider, e.g., the MACRO and Baksan 
experimental spectra, that, as observed in the previous subsection,
are characterized by the same threshold ($E_\mu>1$ GeV)
and thus can be directly compared by superposition.  Both the 
Kolmogorov-Smirnov and the $\chi^2$ tests give a $\ll 1\%$ probability that 
these two distributions are the same  within the {\em statistical\/} errors.  
This suggests that the  experimental systematics might be nonnegligible 
(in some bins, at least),  as indeed confirmed by
the MACRO error analysis  \cite{Ro96}. Unfortunately,  no analogous error 
analysis has been  reported by the Baksan Collaboration.

	The presence of relatively large ``dips and bumps'' in the individual 
experimental spectra of Fig.~\ref{fig:1} poses some practical problems in fitting the 
data with theoretical curves (that, conversely, are rather smooth even in 
the presence of neutrino oscillations). In fact, the oscillation fits to the 
individual spectra  become dominated by the bins that  exhibit the largest 
fluctuations, thus producing rather unstable and unreliable results. 
Therefore, it seems more reasonable to combine first the data from different 
experiments, in order to  ``average out'' their fluctuations, and then to fit
the neutrino oscillation parameters.

\begin{figure}
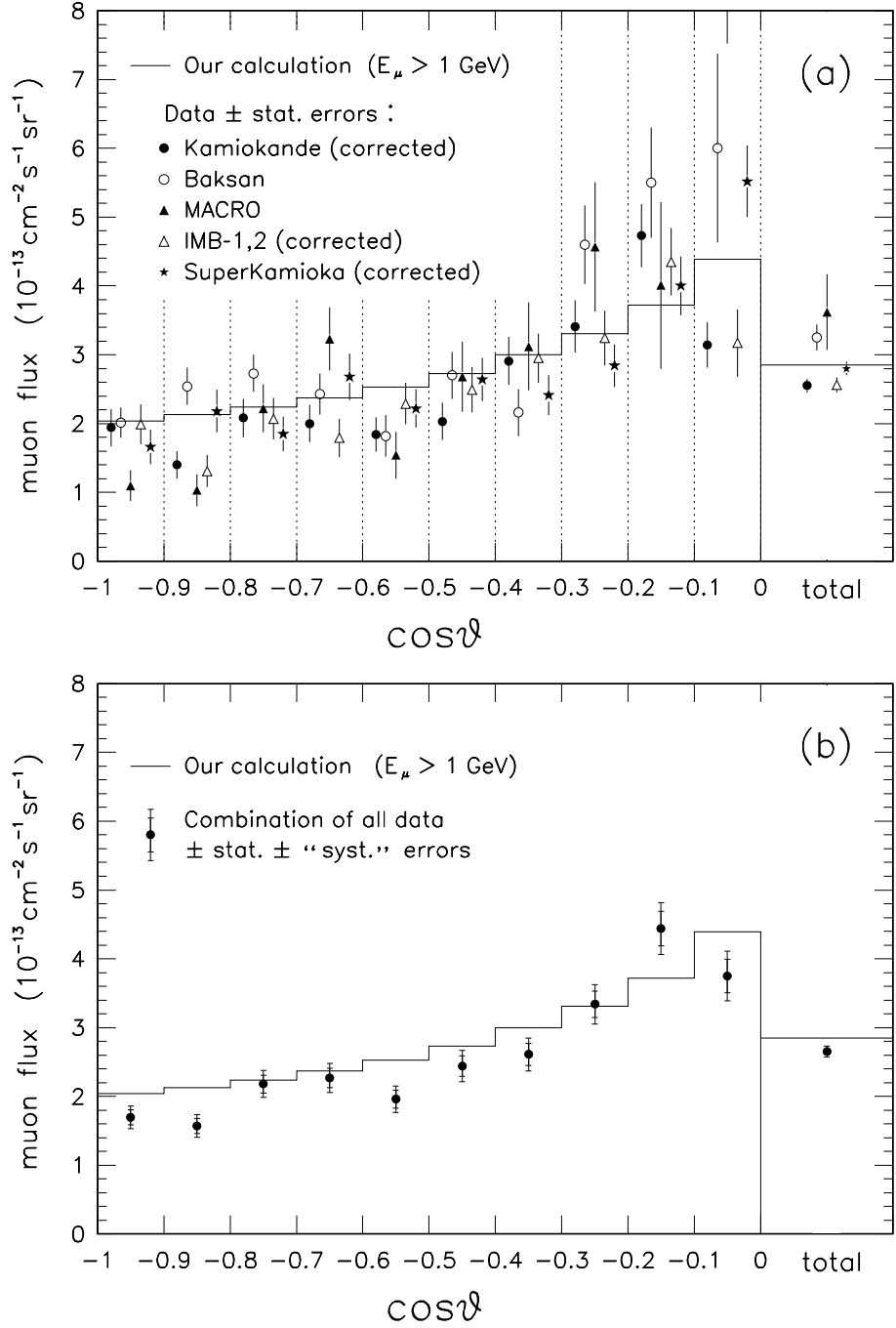

\caption{	Comparison of all upward-going muon spectra, rescaled to a
		common threshold of 1 GeV. (a) Separated data. (b) Combined
		data (as used in our analysis). See the text for details.}
\label{fig:2}
\end{figure}

	We have combined  the five muon flux spectra of Fig.~\ref{fig:1}
into a single spectrum as follows. First,  the experimental distributions have 
been ``rescaled''  to a common 1 GeV threshold, by adding in each bin
the theoretical contribution due to the change in threshold
(no correction is necessary for MACRO and Baksan). The results are shown in 
Fig.~\ref{fig:2}(a), together with the theoretical expectation 
(solid histogram), now
common to all experiments.%
\footnote{
Notice that the spread of the experimental data points in Fig.~\ref{fig:2}(a) 
is significant (a factor of about two in the muon fluxes), suggesting again 
and independently  that large and not well-understood
systematics might affect (at least some of) the experimental results.
}
\ Then the data  have been combined in each bin, using only the known, 
statistical errors. The resulting, combined distribution is shown in 
Fig.~\ref{fig:2}(b) 
(dots with small error bars): it appears smoother than  each of the 
individual parent distributions, and thus  more appropriate for fits.  
A delicate point is the estimate of the (nonnegligible but unknown) 
systematic errors.  The Particle Data Group  recipe for scaling data errors 
as $\sqrt{\chi^2/(N_{\rm DF}-1)}$ \cite{PDBR} would give scaling factors 
between 1 and 2 in the various bins.  Similarly, the  addition of plausible, 
${\cal O}(10\%)$ systematic errors would give total errors $\sim 1.5$ times 
larger than the statistical errors. Guided by these estimates, we have chosen 
to rescale the statistical errors by a uniform factor 1.5 [large error bars 
in Fig.~\ref{fig:2}(b)] in order to obtain the ``total'' uncertainties. 
These ``inflated'' experimental errors have been used  
in all the following fits.

	We stress that any  conceivable combination of the {\em present\/} 
muon data necessarily involves some approximations or
subjective choices, both because spectra with different thresholds cannot be 
compared directly and because no experiment (excepted MACRO) reports
the systematic uncertainties in each bin. For instance, the (somewhat
arbitrary) choice of a common threshold for rescaling the experimental spectra
has some influence on the sensitivity of the fits to low values of the 
neutrino mass differences. Therefore, great care should be taken in 
interpreting the results of {\em any\/} fit to the available muon data,
including those presented in the following Section. Hopefully, in the next 
few years the SuperKamiokande experiment will collect much higher muon 
statistics  and will carefully estimate the systematic uncertainties of the 
flux distribution,  so as to render such drawbacks
gradually less significant in future analyses.

\section{Upward-going Muons and Neutrino Oscillations}

	In this section we report the results of our two-flavor
and three-flavor oscillation analysis of the upward muon spectrum
shown in Fig.~\ref{fig:2}(b). Since this combined spectrum is in reasonable agreement 
with the expectations in the absence of oscillations, exclusion zones
can be drawn in the neutrino oscillation parameter space. This contrasts
with the flavor anomaly observed in contained and semicontained 
atmospheric neutrino events, which suggests nonzero neutrino masses 
and mixings \cite{Fo97}. However, as we will see, positive 
and negative  indications are compatible, within the experimental and 
theoretical uncertainties, in some zones of the parameter space where the 
neutrino mixing is ``intermediate'' (i.e., neither zero nor maximal).

	We use the three-flavor oscillation formalism defined in our 
previous works \cite{Fo96,Fo97,Sc95}. The first two of the three neutrino 
mass eigenstates $(\nu_1,\,\nu_2,\,\nu_3)$ are assumed to be essentially 
degenerate $(m_1\simeq m_2)$, and separated from the third state $\nu_3$ 
by a square mass gap $m^2=|m^2_3-m^2_{1,2}|$. The flavor content of $\nu_3$ 
is parametrized as
$$
\nu_3 = \sin\phi\,\nu_e + \cos\phi\,(\sin\psi\,\nu_\mu+\cos\psi\,\nu_\tau)\ ,
$$
where the mixing angles $\phi$ and $\psi$ range between 0 and $\pi/2$.
The cases $\phi=0$, $\psi=\pi/2$, and $\psi=0$ correspond
to pure two-flavor oscillations in the channels 
$\nu_\mu\leftrightarrow\nu_\tau$,  $\nu_e\leftrightarrow\nu_\mu$, and 
$\nu_e\leftrightarrow\nu_\tau$, respectively (the latter case 
$\nu_e\leftrightarrow\nu_\tau$ is uninteresting for atmospheric
neutrinos and will not be specifically considered). 
As shown in \cite{Fo97,Sc95}, it is useful to chart the parameter space 
through the logarithmic coordinates $(m^2,\,\tan^2\phi,\,\tan^2\psi)$. 
Our results will be presented in representative sections of this space.

	For each value of the neutrino mass and mixing parameters,
we calculate the theoretical muon flux spectrum with the inputs 
described in the caption to Table~\ref{tab:1},  including the matter 
oscillation effect for neutrino paths in the Earth \cite{Fo96,Fo97}. Then we
evaluate the goodness-of-fit to  the experimental distribution in 
Fig.~\ref{fig:2}(b) 
through a $\chi^2$ test. The $\chi^2$ quadratic form includes both the 
uncorrelated  total experimental errors and the correlated theoretical 
uncertainties  of the absolute neutrino fluxes, cross sections,  and 
muon energy losses in the rock.  The $1\sigma$ total theoretical error is
conservatively assumed to be $30\%$, with a bin-by-bin correlation
coefficient equal to $+1$. For both two-flavor and three-flavor 
oscillations the minimum value of $\chi^2$ is searched, and 
C.L. contours are drawn at the values of $\Delta\chi^2=\chi^2-\chi^2_{\rm min}$
appropriate to $N_{\rm DF}=2$ and $N_{\rm DF}=3$, respectively
(see also \cite{Fo97}).  We will combine in different ways
the following data sets,  in order to show their relative ``weight''
in driving the fit: upward-going muons (U) from Kamiokande, IMB-1,2,
Baksan, MACRO, and SuperKamiokande;
sub-GeV (S) $\mu$-like and $e$-like events  from Kamiokande,
IMB, NUSEX, and Frejus \cite{Fo97}; multi-GeV (M) 
$\mu$-like and $e$-like events from Kamiokande \cite{Fu94}, both binned (Mb)
and unbinned (Mu) \cite{Fo97}. The data sets S, Mu, and Mb are analyzed
as in \cite{Fo97}.

\subsection{Two-flavor oscillations}

	Figure~\ref{fig:3} shows the results of our analysis of 
$\nu_\mu\leftrightarrow\nu_\tau$ oscillations. In the first panel (U) only 
the upward-going muon data  are fitted. The symmetry of the C.L.\ contours 
with respect to the vertical axis $\phi=\pi/4$ reflects the absence of matter 
effects in the $\nu_\mu\leftrightarrow\nu_\tau$ channel \cite{Fo97}.
As expected, large neutrino mixing is excluded (for  $m^2\gtrsim 10^{-3}$). 
This negative indication, however, does not rule out the evidence for 
oscillations coming from sub-GeV and multi-GeV binned or unbinned data 
(S+Mb and S+Mu panels).  In fact, the combination of all atmospheric data 
still favors  the oscillation hypothesis, as evidenced in the plots labeled
``S+Mb+U'' (sub-GeV + multi-GeV binned + up-going $\mu$) and
``S+Mu+U'' (sub-GeV + multi-GeV unbinned + up-going $\mu$).

	Notice that the addition of the upward-going muon data 
reinforces the exclusion of large  $\nu_\mu\leftrightarrow\nu_\tau$ mixing 
(required by the negative results of the NUSEX and Frejus experiments
\cite{Fo97}), but does not alter dramatically the oscillation fit to sub-GeV 
and multi-GeV atmospheric neutrino data. In particular, for large $m^2$ the 
``S+Mu+U'' bounds on $\tan^2\phi$ are only slightly narrower than those without upward $\mu$ data
(``S+Mu'' panel). This observation is relevant for those phenomenological
models in which the controversial zenithal anomaly of multi-GeV data is 
discarded, such as the  ``minimum sacrifice'' scenario studied in 
\cite{Ca96,Li97}. We have verified that  the solution to the neutrino 
oscillation evidence proposed in such scenario survives after the inclusion 
of the upward-going muon data, although with a slightly worse $\chi^2$.

	Figure~\ref{fig:4} shows the results of our analysis of 
$\nu_e\leftrightarrow\nu_\mu$ oscillations. The panels are labeled as in 
Fig.~\ref{fig:3}. Notice that the bounds from upward-going muon data  in the 
$\nu_e\leftrightarrow\nu_\mu$ channel are considerably weaker than in the
$\nu_\mu\leftrightarrow\nu_\tau$ channel, and basically exclude a relatively 
small zone at 90\% C.L.\ for large mixing and $m^2\simeq 0.06$--$0.3$ eV$^2$  
(no constraints appear at 99\% C.L.). In fact, in the  
$\nu_e\leftrightarrow\nu_\mu$ channel the $\nu_\mu$ disappearance is 
partially compensated by the conversion of $\nu_e$ (initially present
in the beam) into $\nu_\mu$. The ``suppression'' of the muon disappearance
is also influenced by matter effects, that are responsible of the
left-right asymmetry in the plots (see \cite{Fo97}).  
Also in this case, the bounds from sub-GeV and multi-GeV data (``S+Mu'' 
and ``S+Mb'' panels) are only slightly altered by the inclusion of 
upward-going muon data (``S+Mu+U'' and ``S+Mb+U'' panels, respectively).

	In conclusion, the upward-going muon data tend to disfavor large 
mixing in the $\nu_\mu\leftrightarrow\nu_\tau$ channel and, more weakly, 
in the  $\nu_\mu\leftrightarrow\nu_e$ channel. This indication reinforces
the exclusion of large mixing required by the negative sub-GeV
results from NUSEX and Frejus. However, the indications  for ``intermediate'' 
mixing derived by a global fit to the atmospheric data are not dramatically 
altered by the inclusion of the upward-going muon data. In particular, the
$\nu_\mu\leftrightarrow\nu_\tau$ oscillation mode is still compatible
with the ``minimum sacrifice scenario'' of Refs.~\cite{Ca96,Li97},
which is thus able to accommodate all present neutrino oscillation data,
with the only exception of the non-flat zenith angle distribution
measured in the Kamiokande multi-GeV experiment \cite{Fu94}.

\subsection{Three-flavor oscillations}

	Figure~\ref{fig:5} 
reports the results of our analysis of upward-going muon data
in the three-flavor space $(m^2,\,\tan^2\phi,\,\tan^2\psi)$,  shown as planar 
sections at representative values of $m^2$. We recall that, in each 
$(\tan^2\psi,\,\tan^2\phi)$ panel, the right side corresponds 
(asymptotically) to pure $\nu_\mu\leftrightarrow\nu_e$ oscillations, while 
the lower side correspond (asymptotically) to pure 
$\nu_\mu\leftrightarrow\nu_\tau$ oscillations. The excluded regions in 
Fig.~\ref{fig:5} connect 
continuously  these two subcases and show that also genuine 
three-flavor cases are disfavored by the available upward-going muon data. 
These cases correspond to large neutrino mixings. In particular, the threefold 
maximal mixing scenario \cite{Harr,Giun}, obtained for  
$(\sin^2\psi,\,\sin^2\phi)=(1,\,1/2)$, is  rather strongly disfavored 
by the upward-going muon data.

	The exact shapes of the excluded regions  should be taken with 
some caution, since we have checked that  they are rather sensitive to 
variations in the data and in their statistical treatment; therefore,
new and more precise upward-going muon data from the SuperKamiokande
experiment might alter significantly the bounds shown in Fig.~\ref{fig:5}.
However, the following qualitative features seem to be relatively stable:
1) weaker bounds in the $\nu_\mu\leftrightarrow\nu_e$ channel; and
2) exclusion of large neutrino mixing and, in particular, of
three-fold maximal mixing.

\subsection{Combination of all atmospheric data}

	Figure~\ref{fig:6} 
shows the results of a combined analysis of all the available
sub-GeV, multi-GeV (binned), and upward-going muon data. As 
for Fig.~\ref{fig:5}
we recall that, because of the quality of the available upward-going muon 
data, these fits might change significantly with new, more accurate data. 
Nevertheless, some interesting, qualitative features are seen to emerge:
1) in general, $\nu_e\leftrightarrow\nu_\mu$ oscillations tend to give
a better fit than $\nu_\mu\leftrightarrow\nu_\tau$ oscillations, as a result
of the combined preference of multi-GeV data  \cite{Fo97} and of weaker 
bounds from upward-going muon data for $\nu_e\leftrightarrow\nu_\mu$ 
transitions; and 2) the overall oscillation fit is reasonably good even 
for $m^2$ as small as $\sim 5\times 10^{-4}$ eV$^2$.
These indications suggest that oscillation signals might
be seen better in long-baseline {\em reactor\/} experiments, that can
probe the $\nu_e$ disappearance mode down to $m^2\sim 10^{-3}$ eV$^2$.
If no oscillation signal is found in such experiments, large
horizontal stripes will be excluded in the panels of 
Figs.~\ref{fig:5} and \ref{fig:6} 
\cite{Sc95}.

	In conclusion, both the positive and the negative indications for
flavor oscillations coming from all the available atmospheric neutrino data 
(including upward-going muons) can be made compatible, within the 
uncertainties, at intermediate values of the neutrino mixing angles. In fact, 
although  the positive indications for $\nu$ oscillations favor large mixing, 
the negative results from  upward-going muon experiments (as well as from 
some sub-GeV experiments like Frejus and NUSEX),  tend to drag the fit 
towards small neutrino mixings and to stabilize it at intermediate values. 
Of course,  the overall compatibility of the different data sets  might be 
no longer guaranteed if the total uncertainties were smaller.
Therefore, it is very important that the magnitude of the experimental 
systematic errors in the muon flux distribution (see Sec.~II~B) is clarified
in the running underground experiments (MACRO, Baksan, and
SuperKamiokande).

\section{Summary and Conclusions}

	We have performed an updated analysis of the available upward-going 
muon data (which disfavor neutrino oscillations), and a comparison with the
atmospheric sub-GeV and multi-GeV data (which globally favor neutrino 
oscillations). We have worked out the oscillation
constraints placed by the observed upward-going muon zenithal distribution,
and identified the zones of the two- and three-flavor mass-mixing
parameter space where all the atmospheric data can be reconciled.
In general, the $\nu_\mu\leftrightarrow\nu_e$ oscillation channel
provides a better fit than the $\nu_\mu\leftrightarrow\nu_\tau$
channel, thus privileging searches with long-baseline reactors.
Large neutrino mixing, including threefold maximal mixing, is generally
disfavored. The ``minimum sacrifice'' fit of Refs.~\cite{Ca96,Li97},
that discards, in part, the atmospheric multi-GeV data,
is not significantly altered by the inclusion of the upward-going
muon information.

\acknowledgments

We acknowledge fruitful discussions with
T.\ Montaruli. We thank S.\ Mikheyev and Y.\ Totsuka 
for useful correspondence.
The work of A.M.\ was supported in part by Ministero dell'Universit\`a e 
della Ricerca Scientifica and in part by INFN.


\begin{table}
\squeezetable
\caption{	Detector characteristics, experimental data, and
		theoretical expectations for upward-going muons. Theoretical
		expectations refer to our calculations using Bartol neutrino 
		fluxes \protect\cite{Ag96}, MRS(G) structure functions 
		\protect\cite{MRSG,PDFL}, and Lohmann-Kopp-Voss muon energy 
		losses  in standard rock \protect\cite{Lo85}.
		See the text for details and comments.}
\medskip
\begin{tabular}{cccccccccccccccc}
Experiment								&&
\multicolumn{2}{c}{Kamiokande \cite{Sz96}}				&&
\multicolumn{2}{c}{Baksan \cite{Su96,Mi96}}				&& 
\multicolumn{2}{c}{MACRO \cite{Ro96}}					&&
\multicolumn{2}{c}{IMB-1,2 \cite{Ca91}\tablenotemark[4]}		 		       &&
\multicolumn{2}{c}{SuperKamioka\ \cite{Su97}}				\\ 
\tableline
Depth	(m.w.e.)							&&
\multicolumn{2}{c}{2700  }						&&
\multicolumn{2}{c}{850   }						&& 
\multicolumn{2}{c}{3700  }						&& 
\multicolumn{2}{c}{1570  }						&&
\multicolumn{2}{c}{2700}      \\
Live time (yr)								&&
\multicolumn{2}{c}{$\sim 7$}						&&
\multicolumn{2}{c}{11.94  }						&& 
\multicolumn{2}{c}{3.06}						&& 
\multicolumn{2}{c}{2.53   }						&&    
\multicolumn{2}{c}{0.63}     \\
Technique								&&
\multicolumn{2}{c}{Cherenkov}						&&
\multicolumn{2}{c}{scintillator}					&& 
\multicolumn{2}{c}{scint.+tracking}					&& 
\multicolumn{2}{c}{Cherenkov}						&&
\multicolumn{2}{c}{Cherenkov}						      \\
Dimensions (m$^3$)							&&
\multicolumn{2}{c}{$\pi\,7.8^2\times16.1$}				&&
\multicolumn{2}{c}{$17\times17\times11$ }				&& 
\multicolumn{2}{c}{$12\times77\times9$ }				&& 
\multicolumn{2}{c}{$18\times17\times22.5$}				&&      
\multicolumn{2}{c}{$\pi\,16.9^2\times36.2$}\\
Threshold								&&
\multicolumn{2}{c}{$E_\mu>3$ GeV}					&&
\multicolumn{2}{c}{$E_\mu>1$ GeV}					&& 
\multicolumn{2}{c}{$E_\mu>1$ GeV}					&& 
\multicolumn{2}{c}{$E_\mu\gtrsim 1.8$ GeV}	                        &&     
\multicolumn{2}{c}{$E_\mu\gtrsim 6$ GeV}\\
No.\ of observed $\mu$'s 						&&
\multicolumn{2}{c}{364}	         					&&
\multicolumn{2}{c}{558 }						&& 
\multicolumn{2}{c}{255}			         			&& 
\multicolumn{2}{c}{430 }						&&      
\multicolumn{2}{c}{267} \\
\tableline
$\cos\theta$								&& 
\multicolumn{2}{c}{muon flux\tablenotemark[1]}				&& 
\multicolumn{2}{c}{muon flux\tablenotemark[1]}				&& 
\multicolumn{2}{c}{muon flux\tablenotemark[1]} 				&& 
\multicolumn{2}{c}{muon flux\tablenotemark[1]} 				&& 
\multicolumn{2}{c}{muon flux\tablenotemark[1]} 				      \\ 
bin     								&& 
expt.$\pm\sigma_{\rm stat}$&theo. 					&& 
expt.$\pm\sigma_{\rm stat}$&theo. 					&& 
expt.$\pm\sigma_{\rm stat}$&theo. 					&& 
expt.$\pm\sigma_{\rm stat}$&theo.					&&
expt.$\pm\sigma_{\rm stat}$&theo.					      \\
\tableline
$ [-1,\,-0.9] $&&$
1.44\pm0.27$&1.54&&$2.01\pm0.22$&2.04&&$1.10\pm0.22$&2.04&&$1.72\pm0.29$&1.77
							 &&$0.86\pm0.25$&1.24
                                                                              \\
$[-0.9,\,-0.8]$&&$
0.89\pm0.20$&1.62&&$2.54\pm0.27$&2.13&&$1.03\pm0.23$&2.13&&$1.03\pm0.23$&1.85
							&&$1.35\pm0.31$&1.30\\
$[-0.8,\,-0.7]$&&$
1.54\pm0.28$&1.70&&$2.73\pm0.27$&2.24&&$2.22\pm0.35$&2.24&&$1.78\pm0.30$&1.95
							&&$0.98\pm0.25$&1.37\\
$[-0.7,\,-0.6]$&&$
1.44\pm0.27$&1.81&&$2.43\pm0.30$&2.37&&$3.23\pm0.46$&2.37&&$1.49\pm0.28$&2.07
							&&$1.78\pm0.34$&1.47\\
$[-0.6,\,-0.5]$&&$
1.25\pm0.25$&1.94&&$1.82\pm0.30$&2.53&&$1.54\pm0.34$&2.53&&$1.97\pm0.30$&2.21
							 &&$1.26\pm0.28$&1.57\\
$[-0.5,\,-0.4]$&&$
1.41\pm0.27$&2.11&&$2.70\pm0.34$&2.73&&$2.68\pm0.51$&2.73&&$2.16\pm0.33$&2.40
                                                         &&$1.63\pm0.31$&1.72\\
$[-0.4,\,-0.3]$&&$
2.24\pm0.35$&2.33&&$2.16\pm0.34$&3.00&&$3.12\pm0.64$&3.00&&$2.60\pm0.36$&2.65
							&&$1.32\pm0.29$&1.91\\
$[-0.3,\,-0.2]$&&$
2.69\pm0.38$&2.59&&$4.60\pm0.57$&3.31&&$4.57\pm0.94$&3.31&&$2.86\pm3.93$&2.93
							&&$1.66\pm0.31$&2.13\\
$[-0.2,\,-0.1]$&&$
3.94\pm0.46$&2.93&&$5.50\pm0.80$&3.72&&$4.01\pm1.21$&3.72&&$3.93\pm0.49$&3.30     
                                                        &&$2.71\pm0.43$&2.43\\
$ [-0.1,\,0]  $&&$
2.27\pm0.33$&3.52&&$6.00\pm1.37$&4.39&&$12.7\pm5.18$&4.39&&$2.71\pm0.49$&3.93
                                                        &&$4.06\pm0.52$&2.93\\
\tableline
Total muon flux\tablenotemark[1]
        &&$1.91\pm0.10$\tablenotemark[2]&2.21
        &&$3.25\pm0.19$\tablenotemark[3]&2.85
        &&$3.62\pm0.55$\tablenotemark[3]&2.85
        &&$2.22\pm0.11$\tablenotemark[2]&2.51
	&&$1.76\pm0.10$\tablenotemark[2]&1.81
\end{tabular}
\tablenotetext[1]{\scriptsize 
		Units: $10^{-13}\;{\rm cm}^{-2}\,{\rm s}^{-1}\,{\rm sr}^{-1}$.
		Total $\mu$ flux (bottom row) =
		sum of partial $\mu$ fluxes
		in each $\cos\theta$-bin times the bin width.}
\tablenotetext[2]{\scriptsize Experimental systematic errors not reported in
			      \protect\cite{Sz96,Ca91,Be92,Su97}.}
\tablenotetext[3]{\scriptsize Experimental systematic error of the
			      total flux estimated to be 
			      8\% \protect\cite{Mi96,Ro96}.}
\tablenotetext[4]{\scriptsize IMB-1,2,3 \cite{Be92} (3.6 yr live time)
		did not publish the muon flux distribution. The 617 
		observed muons  were instead
		divided in ``stopping'' (85 events, $1.4\lesssim
		E_\mu\lesssim 2.5$ GeV)
		and ``passing'' (532 events,  $E_\mu\gtrsim 2.5$ GeV).
		The total muon flux reported in \cite{Be92} is 
		$2.22\pm 0.11\times10^{-13}\;
		{\rm cm}^{-2}\,{\rm s}^{-1}\,{\rm sr}^{-1}$ (stat.\ error only). 
		Our theoretical estimates: 102 (stopping) + 645 (passing)
		muons;
		total muon flux = $2.51\times10^{-13}\;
		{\rm cm}^{-2}\,{\rm s}^{-1}\,{\rm sr}^{-1}$.}
\label{tab:1}
\end{table}




\begin{figure}
\caption{	Results of the oscillation fit in the two-flavor oscillation
		channel $\nu_\mu\leftrightarrow\nu_\tau$, shown as C.L.\
		contours in the $(\tan^2\psi,\,m^2)$ plane. The allowed
		regions are marked by stars.}
\label{fig:3}
\end{figure}
\begin{figure}
\caption{	Results of the oscillation fit in the two-flavor oscillation
		channel $\nu_\mu\leftrightarrow\nu_e$, shown as C.L.\
		contours in the $(\tan^2\phi,\,m^2)$ plane. The allowed regions
		are marked by stars.}
\label{fig:4}
\end{figure}
\begin{figure}
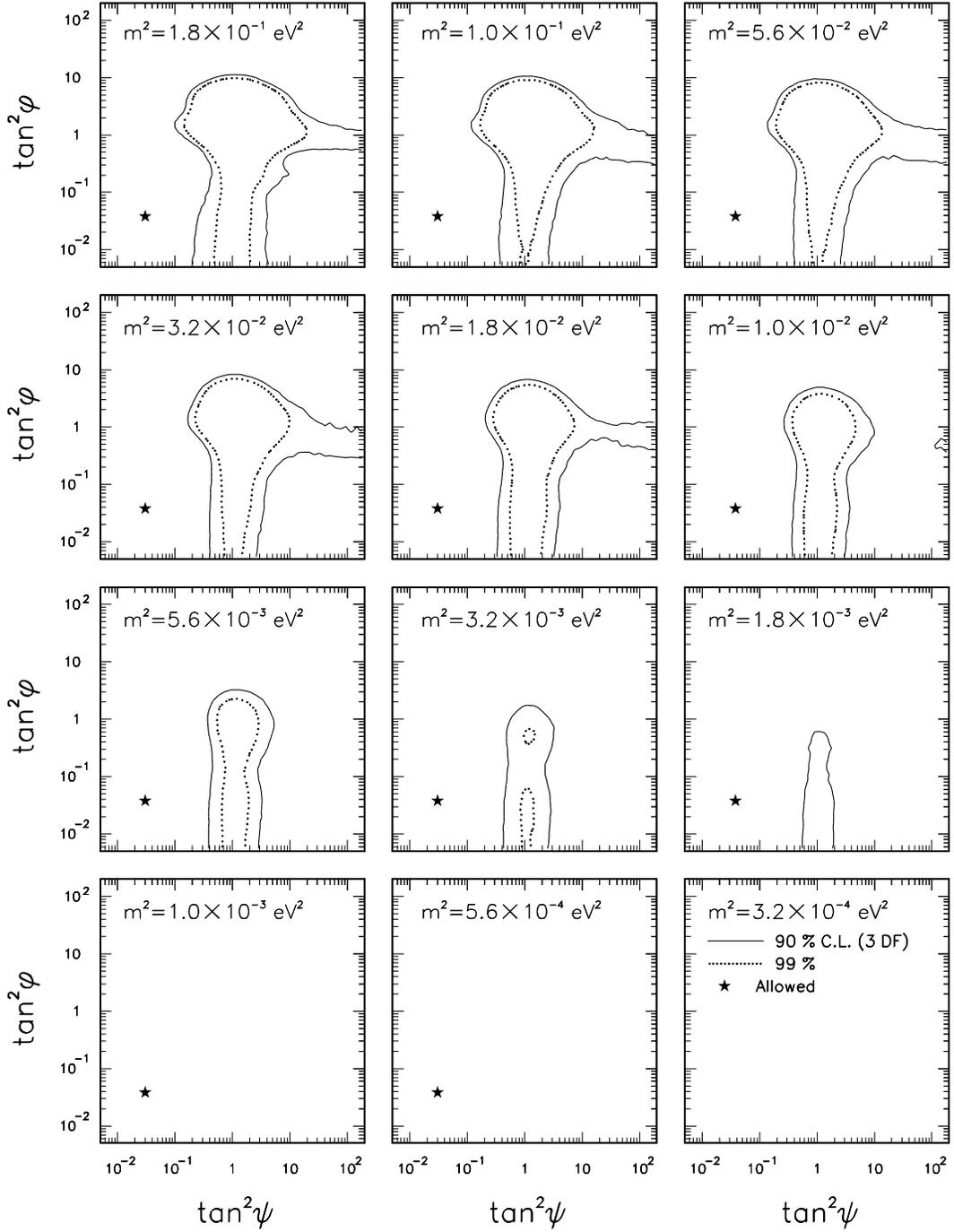

\caption{	Bounds placed by all upward-going muon data in the 
		three-flavor oscillation space
		$(m^2,\,\tan^2\phi,\,\tan^2\psi)$. The allowed regions are
		marked by stars.}
\label{fig:5}
\end{figure}
\begin{figure}
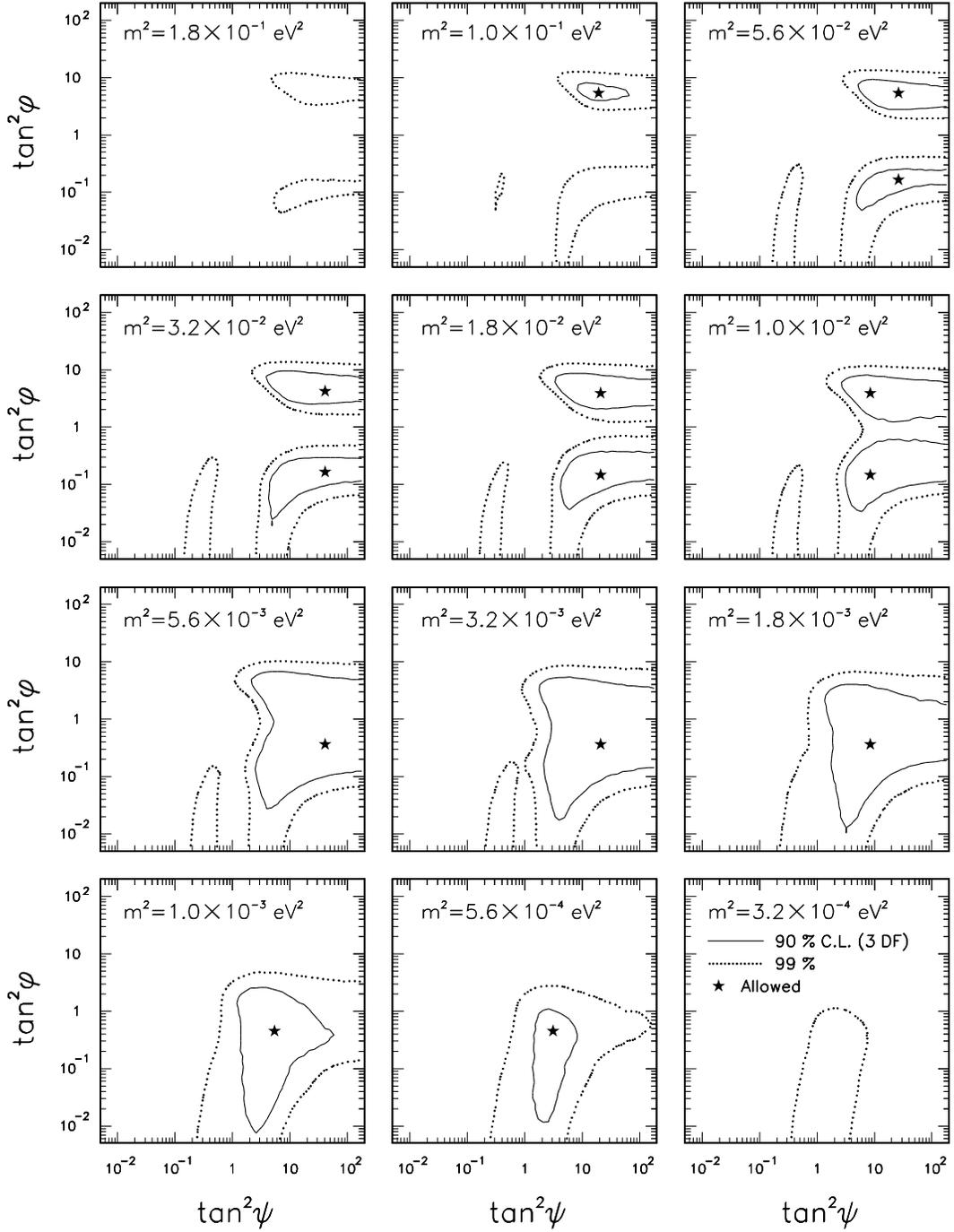

\caption{	As in Fig.~\protect\ref{fig:5}, but with the addition 
		of the atmospheric
		sub-GeV (S) and multi-GeV binned (Mb) data 
		(analyzed as in \protect\cite{Fo97}).
}
\label{fig:6}
\end{figure}


\newcommand{\InsertFigure}[2]{\newpage\begin{center}\mbox{%
\epsfig{bbllx=1.4truecm,bblly=1.3truecm,bburx=19.5truecm,bbury=26.5truecm,%
height=21.truecm,figure=#1}}\end{center}\vspace*{-1.85truecm}%
\parbox[t]{\hsize}{\small\baselineskip=0.5truecm\hskip0.5truecm #2}}

\InsertFigure{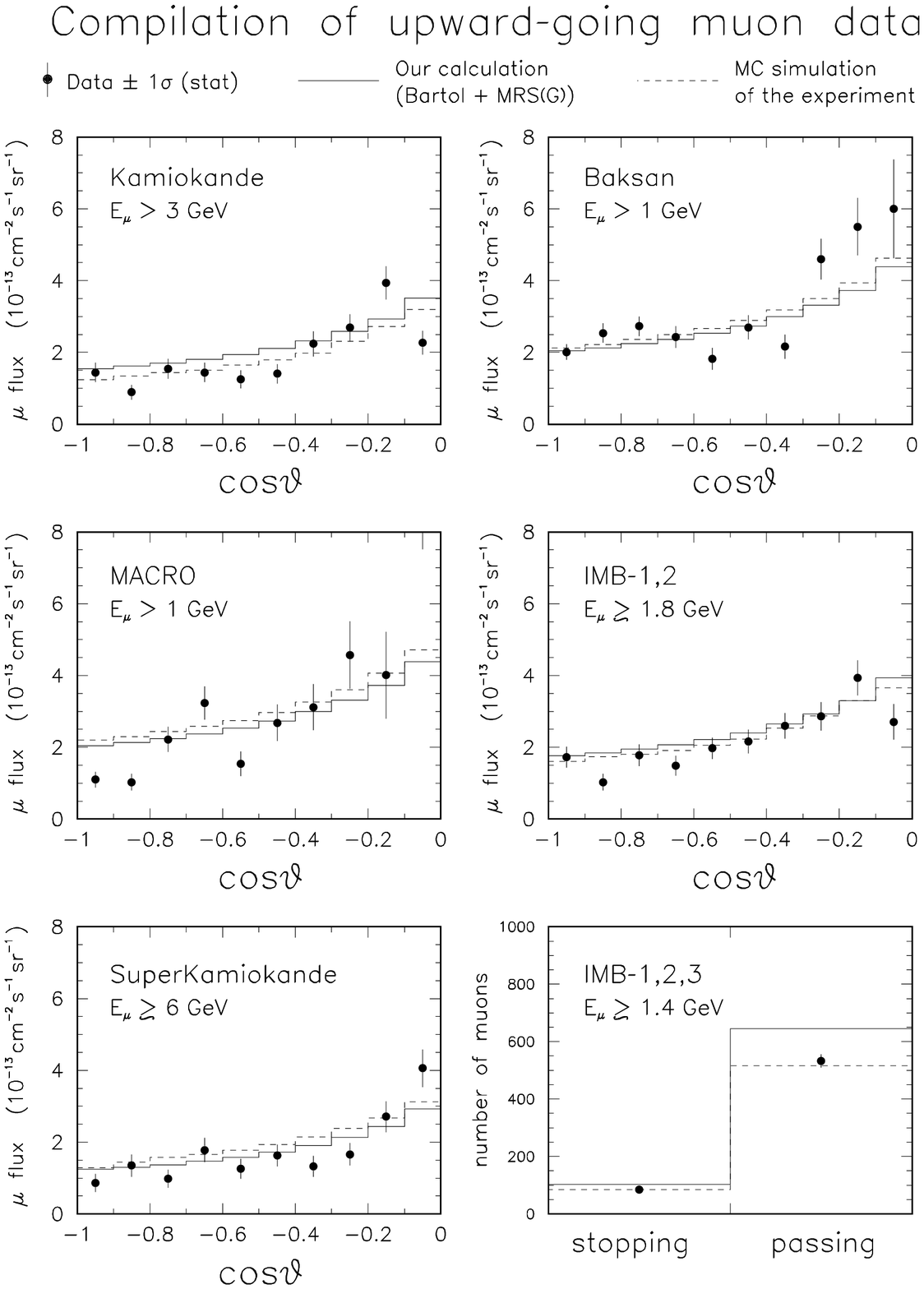}%
{FIG.~1. 	Compilation of upward-going muon data and expectations
		for different experiments: 
		Kamiokande \protect\cite{Sz96},
		Baksan \protect\cite{Mi96,Su96},
		MACRO \protect\cite{Ro96},
		SuperKamiokande \protect\cite{Su97}
		IMB-1,2 \protect\cite{Ca91},
		and IMB-1,2,3 \protect\cite{Be92}. Dots with
		error bars: Experimental data with $1\sigma$
		statistical uncertainties. Solid histograms:
		our calculations. Dashed histograms: Experimental
		simulations. $\theta$ is the zenith angle. See also
		Table~I.
}
\InsertFigure{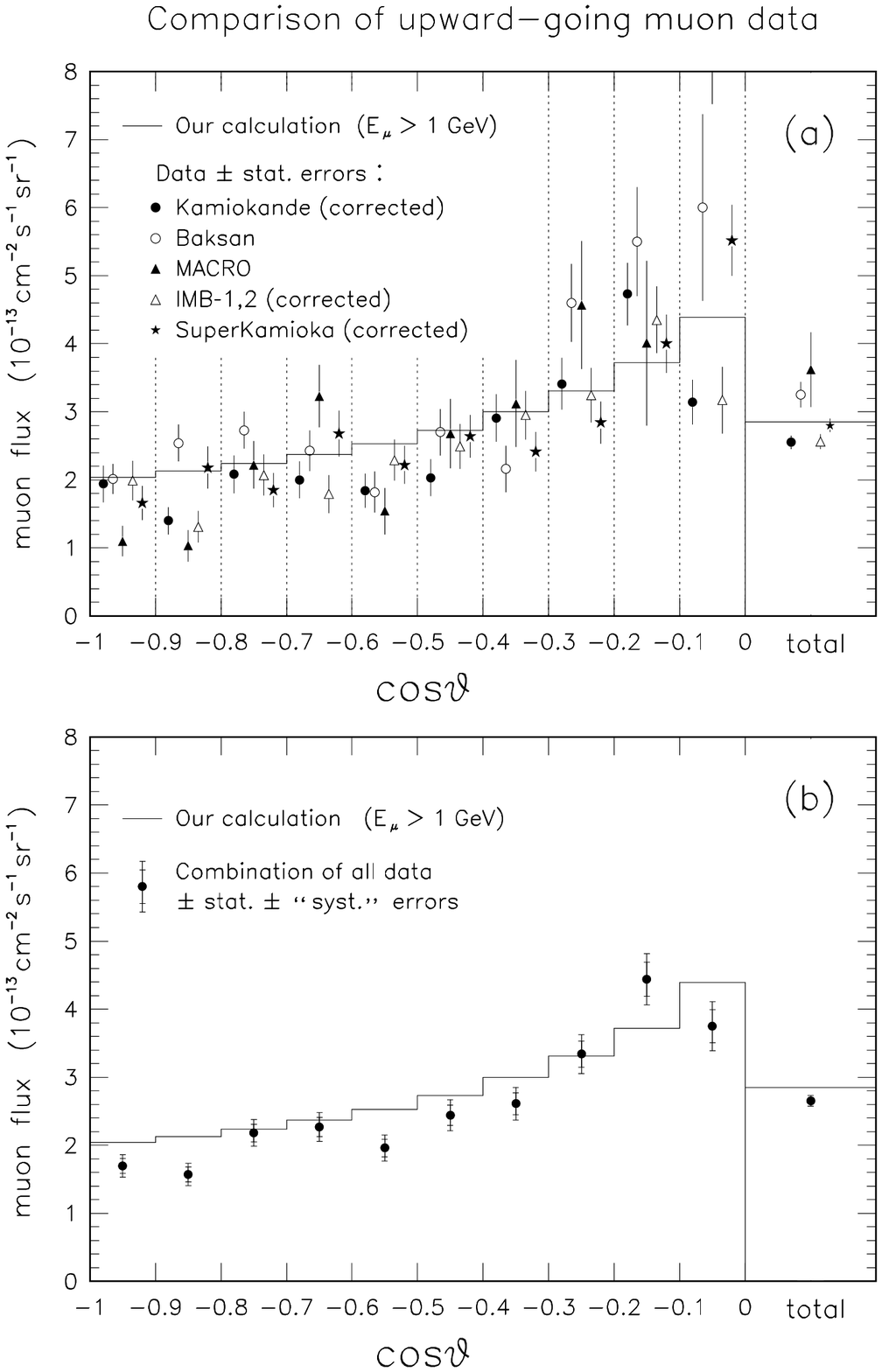}%
{FIG.~2. 	Comparison of all upward-going muon spectra, rescaled to a
		common threshold of 1 GeV. (a) Separated data. (b) Combined
		data (as used in our analysis). See the text for details.
}
\InsertFigure{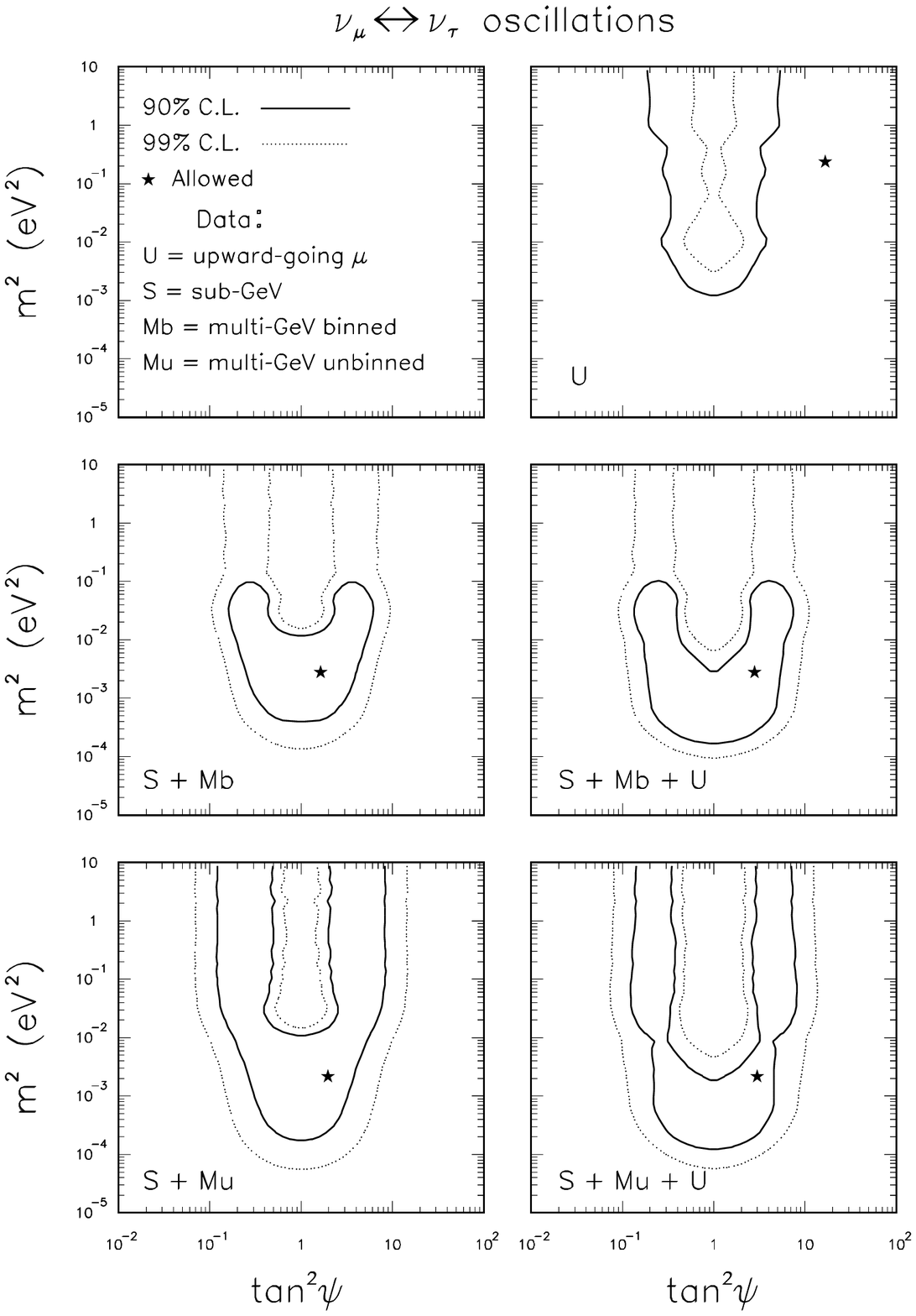}%
{FIG.~3. 	Results of the oscillation fit in the two-flavor oscillation
		channel $\nu_\mu\leftrightarrow\nu_\tau$, shown as C.L.\
		contours in the $(\tan^2\psi,\,m^2)$ plane. The allowed
		regions are marked by stars.
}
\InsertFigure{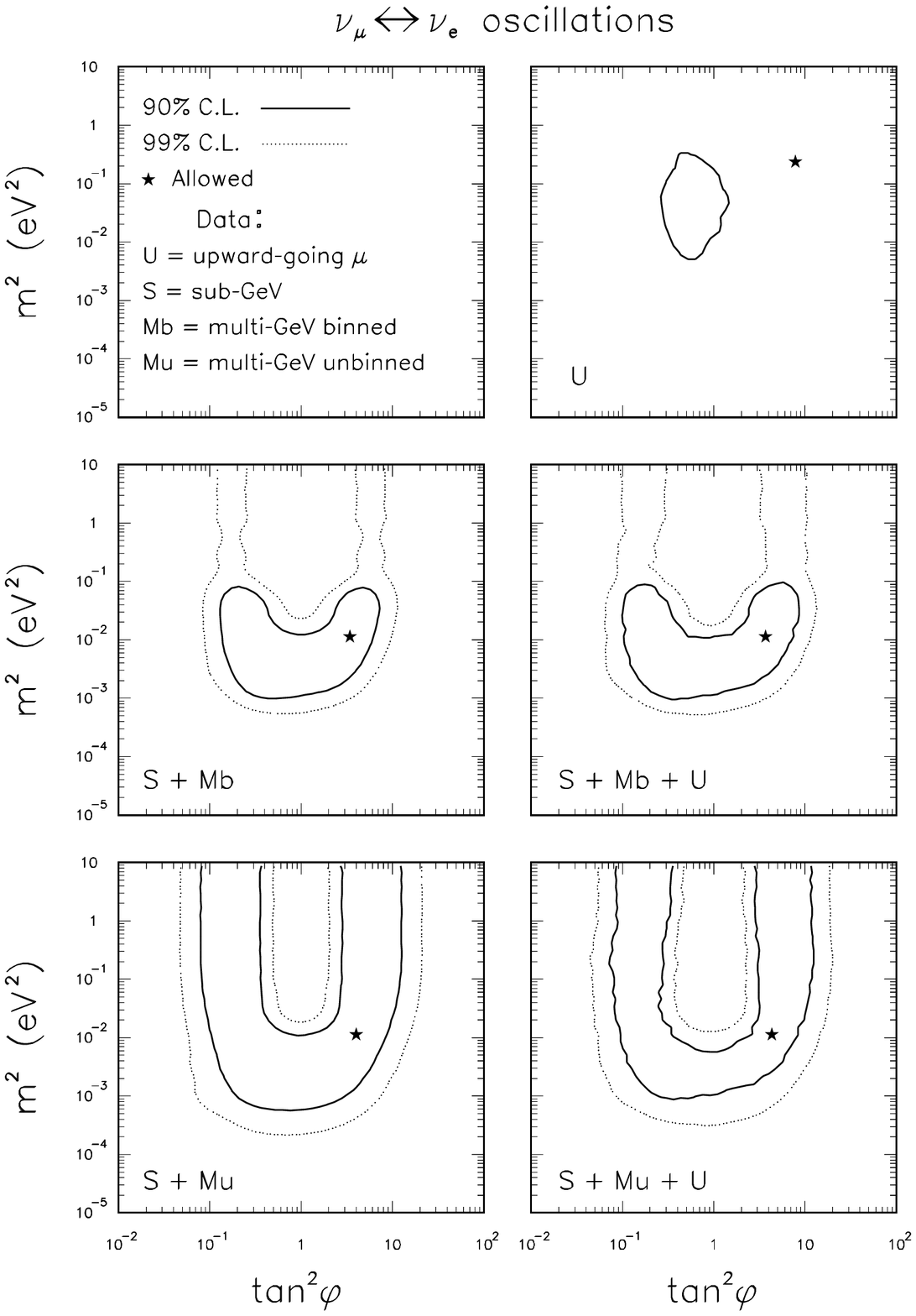}%
{FIG.~4.	Results of the oscillation fit in the two-flavor oscillation
		channel $\nu_\mu\leftrightarrow\nu_e$, shown as C.L.\
		contours in the $(\tan^2\phi,\,m^2)$ plane. The allowed regions
		are marked by stars.
}
\InsertFigure{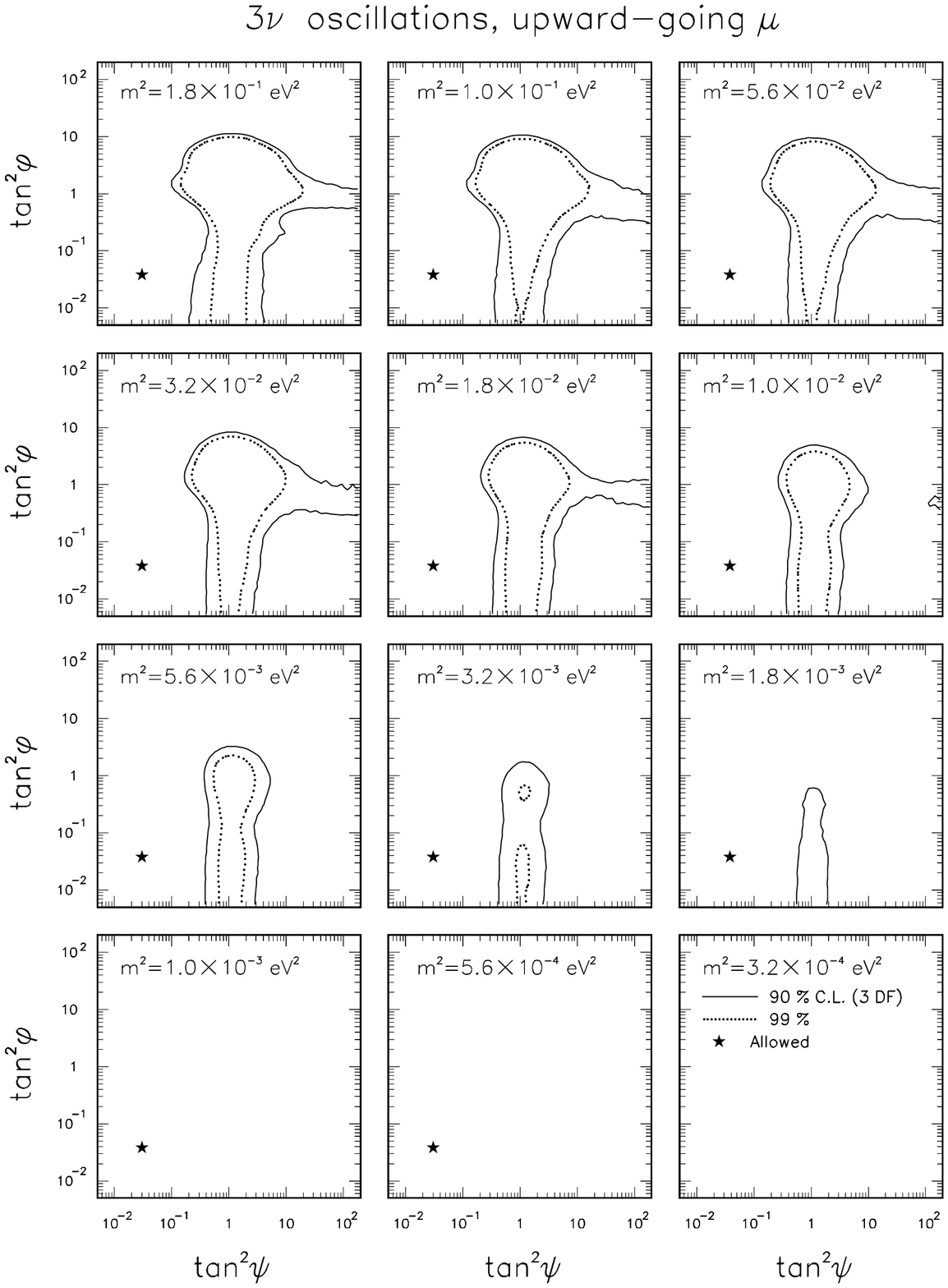}%
{FIG.~5.	Bounds placed by all upward-going muon data in the 
		three-flavor oscillation space
		$(m^2,\,\tan^2\phi,\,\tan^2\psi)$. The allowed regions are
		marked by stars.
}
\InsertFigure{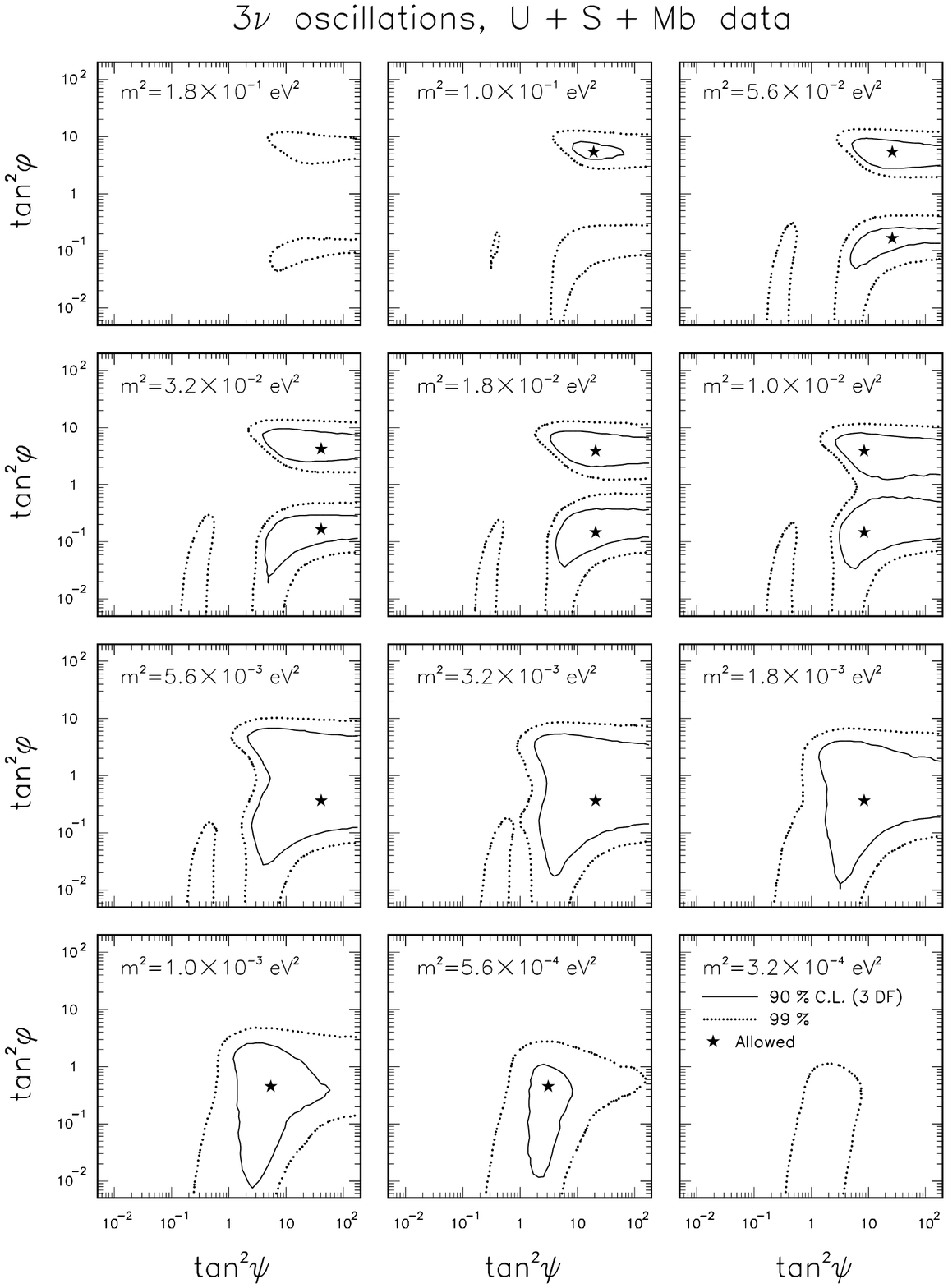}%
{FIG.~6.	As in Fig.~\protect\ref{fig:5}, but with the addition 
		of the atmospheric
		sub-GeV (S) and multi-GeV binned (Mb) data 
		(analyzed as in \cite{Fo97}).
}

\end{document}